\documentclass[pra,twocolumn,showpacs,superscriptaddress]{revtex4-1}
\usepackage[english]{babel}
\usepackage[dvips]{graphicx}
\usepackage{color}
\usepackage{latexsym}
\usepackage{amsmath}
\usepackage{amsthm}
\usepackage{amsfonts}
\usepackage{amssymb}

\begin{document}

\title{Semi-classical formula for quantum tunneling in asymmetric double-well potentials}

\author{G.\ Rastelli}

\affiliation{Universit{\'e} Grenoble 1/CNRS, LPMMC UMR 5493, B.P. 166, 38042 Grenoble, France}

\begin{abstract}
Despite quantum tunneling has been studied since the advent of quantum mechanics, 
the literature appears to contain no simple (textbook) formula for tunneling in 
generic asymmetric double-well potentials. 
In the regime of strong localization, we derive an succinct analytical 
formula based on the WKB semi-classical approach. 
Two different examples of asymmetric potentials are discussed: when the 
two localized levels are degenerate or not.
For the first case, we also discuss a time-dependent problem 
showing quantum Zeno effect.
\end{abstract}


\date{\today}

\maketitle

\section{Introduction}
\label{intro}

Quantum tunneling is of continuing interest in many contemporary areas 
of physics \cite{Kagan-Leggett:1992}. 
The simplest problem can be formulated as a degree of freedom $x$ whose 
potential energy  $V(x)$ has a double-well shape.  
In the classical limit and zero temperature 
one expects that an initial state  prepared in one well is stable. 
Quantum tunneling allows the possibility to escape from one side 
to the other passing under the classically forbidden region.
Canonical examples are an nitrogen atom in ammonia molecule \cite{Cohen-Tannoudji:1997} 
or an electron in a double quantum dot \cite{Burkard:1999}.

However, in recent years, there has be a breakthrough in the experimental study 
of macroscopic quantum tunneling, which is being used to create and study 
"Schr{\"o}dinger-cat" states.
Examples include Bose-Einstein condensates in a double 
trap \cite{Levy:2007,Anker:2005,Albiez:2005,Shin:2005} and 
quantum superconducting circuits based on Josephson junctions 
\cite{Devoret:1992,Makhlin:2001,Chiorescu:2003,Manucharyan:2009,
Simmonds:2004,Cooper:2004,Johnson:2005}. 
Macroscopic quantum tunneling is important to test the validity of the 
quantum mechanics on scales larger than the atomic one \cite{Leggett:2002}.
The investigation of these fundamental issues will be also useful for advanced  
technological applications, such as the development of devices for 
quantum information processing \cite{Ladd:2010}.

These experiments involve a system tunneling from one macroscopic state to another.
Despite the complexity of this process, it is often remarkably well described by
the physics of a single particle in a double-well potential in which 
the variable $x$ corresponds to a collective macroscopic variable. 

The double-well needs not be symmetric, and in many experiments the asymmetry 
can be changed by modifying externally some tunable parameters 
(see, for instance, Refs.\cite{Simmonds:2004,Cooper:2004,Johnson:2005},   
Ref.\cite{Levy:2007} and references therein).

One can formulate the problem as a potential $V_{\eta}(x)$ whose shape depends 
on a dimensionless parameter $\eta$ that quantifies the asymmetry, 
i.e. $\eta=0$ corresponding to the symmetric case. 
In the limit of high energy barrier, i.e. $V_0 \gg \hbar \omega$ where 
$V_0$ is the energy scale of the barrier and $\omega$ the typical 
harmonic frequency in the wells, the low-energy physics reduces to the 
standard two-level system: $\varepsilon_L(\eta),\varepsilon_R(\eta)$ are the energies  
of two localized states coupled by quantum tunneling with amplitude $\nu(\eta)$.

The last parameter $\nu(\eta)$ has to be determined from the given double-well potential. 
Remarkably, the literature appears to contain no simple (textbook) formula for tunneling 
in generic asymmetric double-well potentials 
\cite{Tomsovic:1998,Takagi:2002,Razavy:2003,Ankerhold:2007,Miyazaki:2007}.
Few exceptions are the works in Refs.\cite{Dekker:1987,Schmidt:1991,Benderskii:1999,Johnson:2005,Konwent:1998}
in which general methods based on sophisticated techniques are discussed but 
useful analytical formulas are presented only for specific shapes of the potential.

\begin{figure}[htbp]
\begin{center}
\includegraphics[scale=0.24,angle=270.]{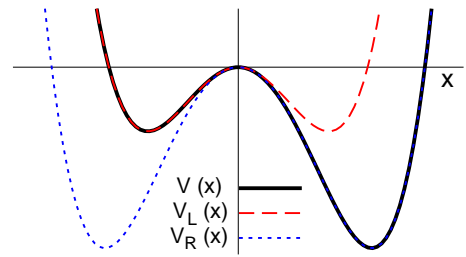}
\end{center}
\caption{(Color on line) The asymmetric potential  $V(x,\eta)$, the solid (black) line, 
is conceived as the result of merging  two symmetric double-well potentials: 
 $V_{\eta}(x) \equiv V_L(x,\eta)$ for $x<0$, the dashed (red) line,
whereas $V_{\eta}(x) \equiv V_R(x,\eta)$ for $x>0$, 
the dotted (blue) line.}
\label{fig:results}
\end{figure}

In this work we revisit the problem. 
By using the standard WKB-approach, we demonstrate that it is possible  to express  
the amplitude $\nu(\eta)$ as the simple formula:   
\begin{eqnarray}
\nu \left( \eta \right) &=& A\left( \eta \right) \sqrt{\nu_{L}(\eta) \nu_{R}(\eta)} \, , \label{eq:result} \\
A\left( \eta \right)  &=& \!\!\!
\frac{1}{2} \!\!
\left[\!
{\left( \frac{V(0)- \varepsilon_L(\eta)}{V(0)- \varepsilon_R(\eta)}\right)}^{\frac{1}{4}}
\!\!+\! 
{\left( \frac{V(0)- \varepsilon_R(\eta)}{V(0)- \varepsilon_L(\eta)}\right)}^{\frac{1}{4}}
\!\right] \! \! , \label{eq:result_1}
\end{eqnarray}
where $\nu_{L}(\eta)$ and $\nu_{R}(\eta)$ are the tunneling amplitudes associated to 
 two {\em symmetric} double-well potentials: $V_{L}(x,\eta)=V_{L}(-x,\eta)$ and 
$V_{R}(x,\eta)=V_{R}(-x,\eta)$.
As shown in Fig.~\ref{fig:results}, they are defined by the equations 
$V_{\eta}(x)\equiv V_{L}(x,\eta)$ for $x<0$ and 
$V_{\eta}(x)\equiv V_{R}(x,\eta)$ for $x>0$, Fig.~\ref{fig:results}. 
$V_{\eta}(0)$ is the maximum of the potential at $x=0$.
%
%
%
%
%
%
%
%
%
%
%
%
%

Now $\nu_L(\eta)$ and $\nu_R(\eta)$ can be easily obtained by using the well-known 
analytic formula for symmetric double-wells (see Refs.\cite{Coleman:1977,Kleinert:1995,Garg}) 
reported in Eqs.~(\ref{eq:nu_s_instanton}),(\ref{eq:C_instanton}) 
of this article for completeness.

Hence the analytical and succinct formula Eqs.~(\ref{eq:result}),(\ref{eq:result_1}) 
together with  the Eq.~(\ref{eq:nu_s_instanton}) allow a direct calculation 
of the tunneling amplitude in an asymmetric double-well.  
To the best of our knowledge, this general formulation  has never been proposed although  
it can be of great use to experimentalists in quantifying their results. 

\begin{figure}[htbp]
\includegraphics[scale=0.17,angle=270.]{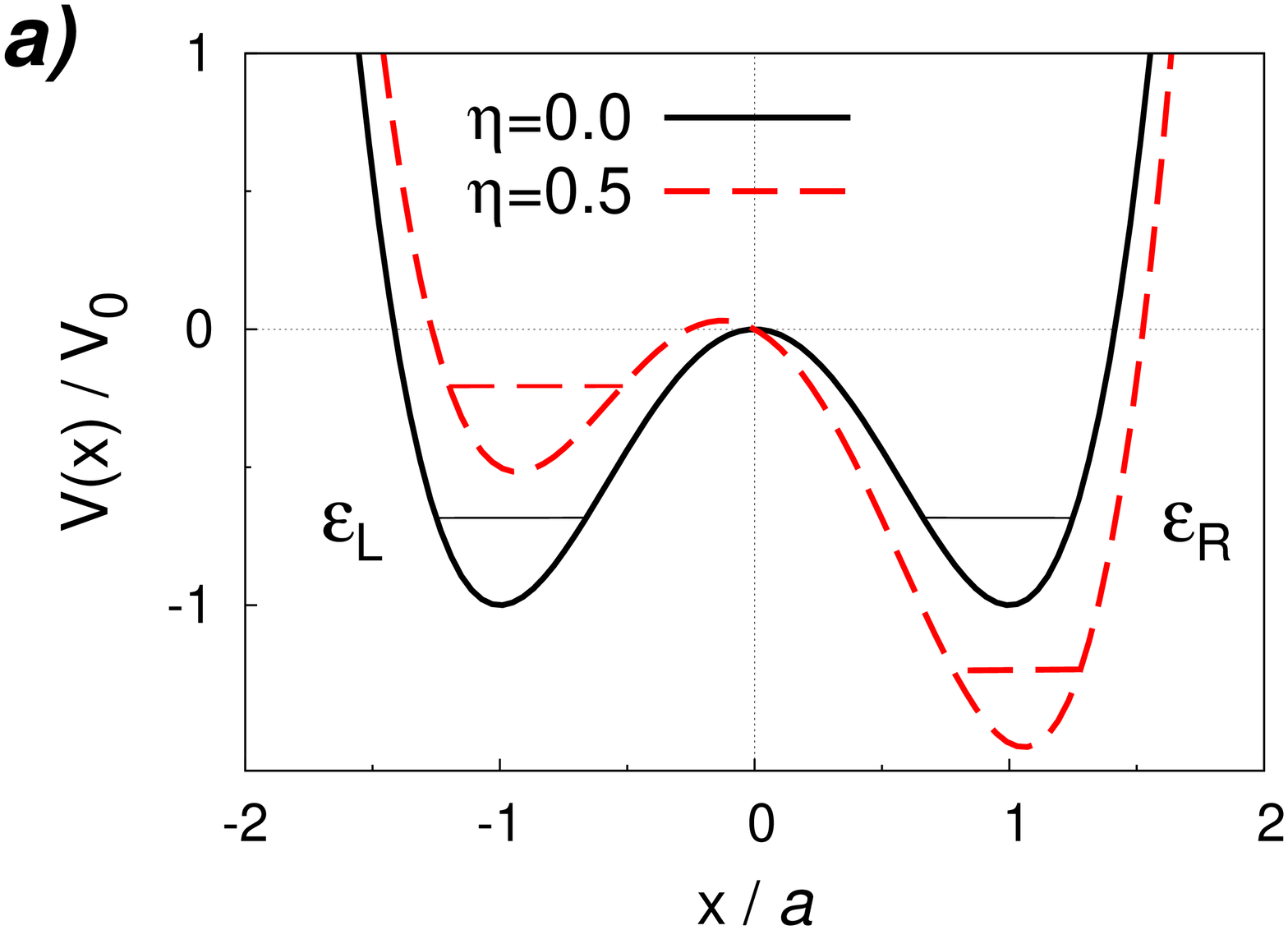}\includegraphics[scale=0.17,angle=270.]{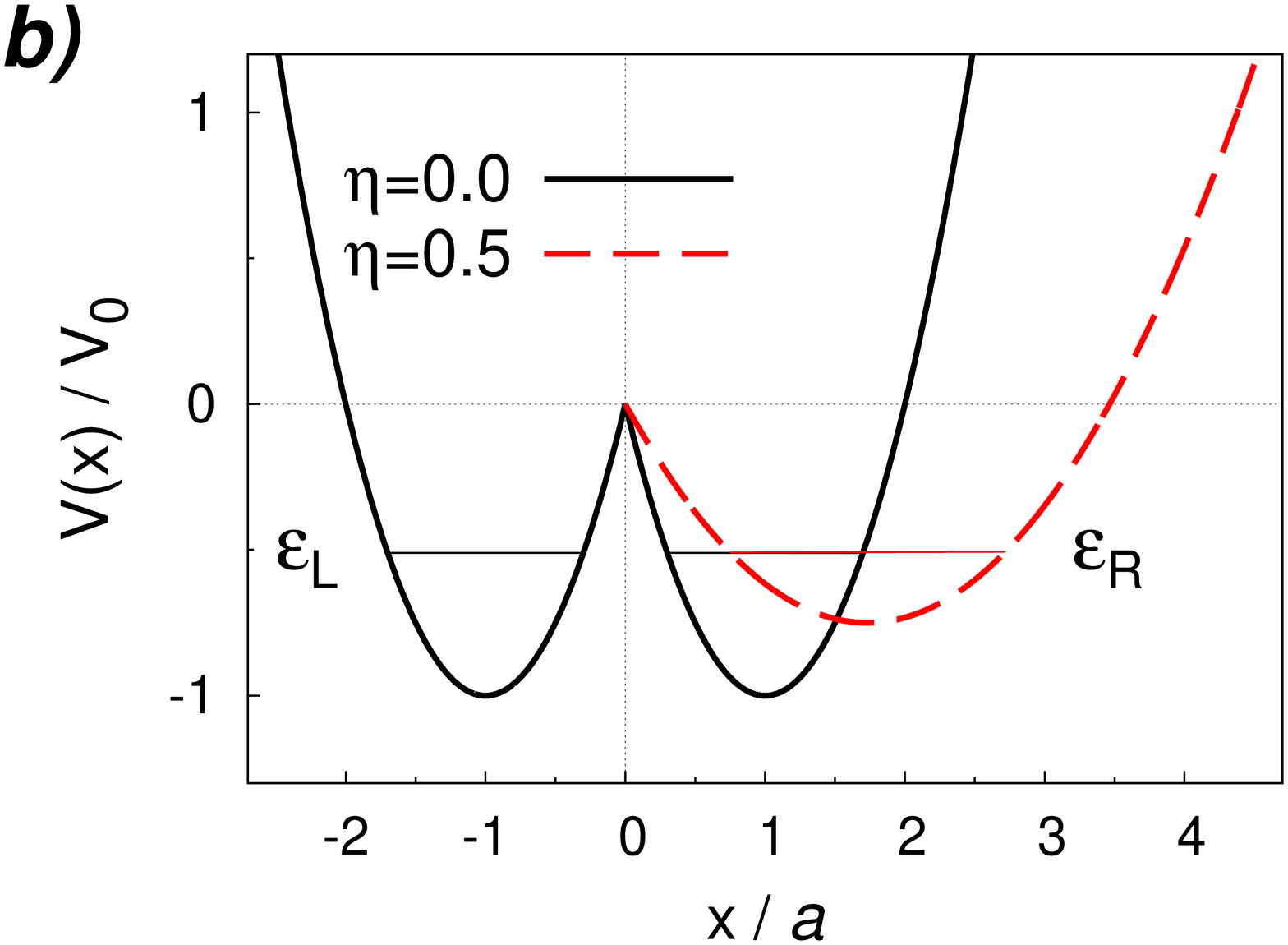}
\caption{(Color on line) Two examples of asymmetric double-well potential 
at different values of $\eta$: $\eta=0$ (symmetric) full lines, 
$\eta=0.5$ dashed (red) lines.  
(a) The quartic potential Eq.~(\ref{eq:V_4}). 
(b) The parabolic potential  of Eqs.~(\ref{eq:V_asym_1}),(\ref{eq:V_asym_2}),(\ref{eq:V_asym_3}) 
with $\varepsilon_{L}=\varepsilon_{R}=-V_0/2$.  
The horizontal lines represent schematically the two energies $\varepsilon_{L}(\eta)$ 
and $\varepsilon_{R}(\eta)$.}
\label{fig:asymmetry} 
\end{figure}

The rest of the paper is organized as follows. 
In Sec.~\ref{sec:TwoLevels} we introduce the two-level system and 
we recall the equivalence of the two semi-classical methods, i.e. the 
WKB approximation and the instanton technique. 
Then, by using the simpler WKB method, we derive the formula 
Eqs.~(\ref{eq:result}),(\ref{eq:result_1})  in Sec.~\ref{sec:demonstration}. 
As an example of application, we discuss two cases in Sec.~\ref{sec:Applications}.
In the first one we consider an asymmetric double-well potential 
in which the degeneracy is removed as the asymmetry is introduced: 
the bias quartic potential with $\varepsilon_L(\eta) \neq \varepsilon_R(\eta)$ for $\eta \neq 0$ 
(see Fig.~\ref{fig:asymmetry}a).
In the second case, we consider a particular situation in which 
the asymmetry of the potential is introduced without removing 
the degeneracy $\varepsilon_L(\eta)=\varepsilon_R(\eta)=\varepsilon$ 
(see Fig.~\ref{fig:asymmetry}b). 
Now the role of the asymmetry in the tunneling dynamics 
appears in a clear-cut way as the Rabi frequency is directly related 
to the tunneling amplitude   $\hbar \Omega = 2\nu(\eta)$. 
For this second case, we also discuss a time-dependent problem  corresponding 
to a special case of the quantum Zeno effect \cite{Misra:1977,Smerzi:2006} in 
an asymmetric double-well potential in which the asymmetry varies with time $\eta(t)$.

The appendix contains comparisons between the exact numerical results and the 
semi-classical formula for the case of the two potentials discussed in this paper. 
%
%
%
%
%
%
%
%
%
%
%
%

\section{Model and approximations}
\label{sec:TwoLevels}

\subsection{Shape of the potential}
 
For the potential $V(x)$, we assume that 
its  maximum is set at the origin  $V(0)=0$, the minima at   
left $x=a_L <0$ and at  right $x=a_R>0$, and that the potential increases 
in the limit  $\left| x \right| \rightarrow \infty$ so that 
the eigenstates $\left\{ \psi_n(x) \right\}$ decay far away the origin 
and the energy spectrum $\left\{ E_n \right\}$ is discrete. 

Here we focus on the low-energy regime of tunneling.
Then, a priori, only a few eigenstates are involved in the dynamics.
Specifically, one can restrict to the two lowest energy states $E_0,E_1$: 
the ground state $\psi_{0}(x)$  and the first excited state $\psi_1(x)$.

\subsection{Two-level effective model}

Around the two minima $s=L,R$, whose local harmonic frequencies are $\omega_{s}$, 
it is possible to solve locally the Schr{\"o}dinger equation to obtain 
two localized states $\phi_L(x)$ and $\phi_R(x)$.
They are well localized into the wells over a range of order 
$\sigma_{s} = {[ \hbar/(2m\omega_{s}) ]}^{1/2}$  ($m$ is the particle's mass)  
when the heights associated 
to the left and right energy barrier are large compared  to the kinetic energies 
of localization
\begin{equation}
\label{eq:condition}
\left[ V(0) - V(a_s) \right] > \frac{\hbar^2}{2m\sigma_{s}^2} \sim \hbar \omega_s \, .
\end{equation}
However $\phi_L(x)$ and $\phi_R(x)$ also decay inside the barrier as we explain below. 

Their energies $\varepsilon_{s} = V(a_s) + \hbar \omega_{s} /2$   are 
assumed to be close to the energy range spanned by $E_0,E_1$. 
We assume that the asymmetry is not too strong to yield resonance 
between the first (approximate) excited state in one well and the 
(approximate) ground state of the opposite well \cite{Schmidt:1991}.
Increasing the asymmetry, the two-level description is no more valid as  
the resonant condition is approached, for instance when   
$\varepsilon_L(\eta) \simeq \varepsilon_R(\eta) + \hbar\omega_R(\eta)$.

In the regime defined by Eq.~(\ref{eq:condition}), 
the linear combination of the two wave functions $\phi_L(x)$ and $\phi_R(x)$ 
is to be the best choice as a (real)  two-terms expansion  to approximate 
the exact ground state and the first excited state.
The most general linear combination reads 
\begin{eqnarray}
\psi_{0}(x) &=& \cos(\theta) \phi_L(x) - \sin(\theta) \phi_R(x) \label{eq:psi_0} \, ,\\
\psi_{1}(x) &=& \sin(\theta) \phi_L(x) + \cos(\theta) \phi_R(x) \label{eq:psi_1} \, .
\end{eqnarray}
Eqs.~(\ref{eq:psi_0}),(\ref{eq:psi_1}) provide the condition of orthonormality between 
$\psi_{0}$ and $\psi_{1}$ assuming that $\phi_L(x)$ and $\phi_R(x)$ are themselves orthogonal 
and normalized.
The above discussion  motivates the introduction  
of the standard two-level Hamiltonian 
\begin{equation}
\label{eq:H_TLS}
\hat{H} = 
\sum_{s=L,R} \varepsilon_s(\eta) \left| s \right> \left< s \right|
- \nu(\eta)  
\sum_{s \neq s'} \left| s \right> \left< s' \right| \, ,
\end{equation}
which describes quantum tunneling between the two 
localized  states in the left well and in the right one and 
$\nu(\eta)$ is the amplitude for tunneling. 
The full Hilbert space is thus spanned by only two states. 
Solving the equation for the two eigenstates  $\psi_{0}$ and $\psi_{1}$ of the $2\times2$ matrix 
Eq.~(\ref{eq:H_TLS}),  one 
  obtains the relation between the angle $\theta$ appearing in Eqs.~(\ref{eq:psi_0}),(\ref{eq:psi_1}) 
and the parameters $\varepsilon_s,\nu$
\begin{equation}
\sin\left( 2 \theta \right) =  
\sqrt{\frac{\nu^2(\eta)}{\nu^2(\eta) + {\left[ \Delta\varepsilon(\eta)/2 \right]}^2}  } \, ,
\end{equation}
where $\Delta\varepsilon(\eta)=\varepsilon_L(\eta)-\varepsilon_R(\eta)$.
For $\Delta\varepsilon = 0$  the rotating angle is  $\theta =\pi/4$.
Then the general Hamiltonian for two-level system with degeneracy Eq.~(\ref{eq:H_TLS}) 
has always eigenstates corresponding to symmetric or antisymmetric linear 
combinations of the two states $L$ and $R$. 
It is worthwhile to stress that this result holds even if  the left and right wave functions 
$\phi_L(x)$ and $\phi_R(x)$  are generally different for an arbitrary asymmetric potential.
On the other hand, the asymmetry affects the tunneling amplitude $\nu(\eta)$ 
as we explain below.

Given the parameters entering the effective Hamiltonian Eq.~(\ref{eq:H_TLS}), 
any physical quantities   can be evaluated. 
For instance, solving the time-dependent Schr{\"o}dinger equation   and assuming 
$\left| \psi(t=0) \right> = \left| L \right>$ at the initial time, 
we obtain $P_{R}(t)$, the probability 
to have the system at right and at the time $t$. 
With the boundary condition $P_{R}(0)=0$, its derivative 
satisfies the following equation 
\begin{equation}
\label{eq:evolution_Pr}
\frac{dP_{R}(t)}{dt} = 
\frac{\nu}{\hbar} 
\sin(2 \theta) 
\sin \left( \Omega t \right) \, ,
\end{equation}
where the Rabi oscillation frequency is $\hbar \Omega = E_1-E_0$
and $E_0,E_1 = (\Delta\varepsilon(\eta)/2) \mp {[  {( \Delta\varepsilon(\eta)/2)}^2 + \nu^2(\eta) ]}^{1/2}$.

\subsection{WKB approximation and instanton technique for symmetric double-well}

To obtain the amplitude $\nu$ for quantum tunneling in the semi-classical 
regime defined by Eq.~(\ref{eq:condition}), the well-known  approaches 
are the instanton technique and the WKB approximation.

The WKB method is based on the Lifshitz-Herring formula \cite{Herring:1962,Landau:1977}.
For a symmetric double-well, it reads 
\begin{equation}
\label{eq:nu_definition_sym}
\nu = \frac{\hbar^2}{m} {\left( \phi \frac{d\phi}{dx}  \right)}_{x=0}  \, , 
\end{equation}
in which $\phi(x)$ represents the (approximated) localized solution in the 
double-well potential $(\phi=\phi_L=\phi_R)$.

In the past the discrepancy of the results given by the two methods has been extensively discussed. 
For different potentials, they resulted 
in a difference by the well-known factor ${(e/\pi)}^{1/2}$ \cite{Gildener:1977,Neuberger:1978}.
For a symmetric potential, Garg showed that this difference is not a failure of the 
formula Eq.~(\ref{eq:nu_definition_sym}) but it was related  to the choice of 
the wave function $\phi$ approximating the local solution \cite{Garg}. 
Introducing $\tilde{x}$ as the classical turning point where the local 
energy $\varepsilon=\varepsilon_L=\varepsilon_R$  intersects the potential barrier 
$\varepsilon=V(\tilde{x})$, the corrected WKB solution inside the barrier $(0<x<\tilde{x})$ 
for the symmetric double-well reads  \cite{Garg,Catelani}
\begin{equation}
\label{eq:WKB_solution_2} 
\phi(x) \!\! = \!\! \sqrt{ \frac{m\omega}{2\pi e k(x)} } \,  
 e^{-\frac{1}{\hbar} \int^{\tilde{x}}_{x} \!\! dx'k(x')}  , 
\, \, 
k(x) = \sqrt{ 2m \left(V(x)-\varepsilon\right) }
 \, , 
\end{equation}
where $k(x)$ is now the inverse  of the local penetration length.
As $dV(x)/dx=0$ at the maximum at $x=0$, we have the relation 
\begin{equation}
\label{eq:derivative_0}
{\left.\frac{d\phi(x)}{dx}\right|}_{x=0} = \frac{1}{\hbar} k(0) \phi(0) \, .
\end{equation}
When the expression Eq.~(\ref{eq:WKB_solution_2}) is inserted in Eq.~(\ref{eq:nu_definition_sym}), 
we obtain the result 
\begin{equation}
\label{eq:t_WKB}
\nu =  \frac{\hbar}{m} k(0) \phi^2(0) \, .
\end{equation}
As discussed in  Ref.\cite{Garg}, expanding Eqs.~(\ref{eq:WKB_solution_2}),(\ref{eq:t_WKB})
around the singular 
point $\tilde{x}$,  it is possible to recover exactly the instanton solution 
\cite{Coleman:1977,Kleinert:1995} in which we have the exponential integrals extending from the 
first $x=-a$ to the second minimum $x=+a$. 
For the sake of completeness, we recall here the result:
\begin{equation}
\label{eq:nu_s_instanton}
\nu =  \hbar \omega \sqrt{\left( \frac{m \omega a^2}{\pi \hbar}\right)} 
e^{C}
e^{-
\frac{1}{\hbar}
\int^{+a}_{-a} \! dx \, \sqrt{2m\left[V(x)-V(a)\right]}
} \, , 
\end{equation}
where  the numerical prefactor  $C$ is 
\begin{equation}
\label{eq:C_instanton} 
C = \int^{a}_0 \! dx \left(  \frac{m\omega}{\sqrt{2 m [V(x)-V(a)]}} - \frac{1}{a-x} \right) \, .
\end{equation}

\subsection{Discussion}

Despite their equivalence for the formula of the quantum tunneling amplitude, 
the instanton technique remains particularly advantageous 
when the particle is coupled to an external environment \cite{Leggett:1987}. 
Then the quantum dissipative dynamics can be formulated in terms of the path integral 
  in which the relevant object is the Euclidean action of the full system 
(and not the wave functions) \cite{Radosz:2006}. 
In the presence of  coupling with a dissipative external bath, a particle moving 
in an asymmetric double-well potential can even  relax towards its minimal configuration 
\cite{Leggett:1987,Grabert:1985}.

The coupling with the environment is weak as long as $\gamma a^2/\hbar \ll 1$ 
where $\gamma$ is the linear friction coefficient \cite{Grabert:1985}.
When the coherent quantum dynamics of the particle in the potential 
is of interest (the decoherence time is much greater than the Rabi frequency), 
the effective two-level model holds and the tunneling amplitude is an 
intrinsic quantity which characterizes the quantum isolated system.  

As Benderskii et al. demonstrated that the two semi-classical methods are equivalent 
even for asymmetric double-well potentials \cite{Benderskii:1999},  
one can choose to use the simpler method to tackle the problem, i.e. the WKB approach.

\section{Semi-classical formula for asymmetric potentials}
\label{sec:demonstration} 

We now formulate the WKB-approach for an arbitrary asymmetric potential.
The starting point is the conservation law for the probability density  
$\rho(x,t) = {\left| \Psi(x,t) \right|}^2$
\begin{equation}
\label{eq:conservation}
\frac{d \rho(x,t)}{dt} = -  \frac{d J(x,t)}{dx} \, ,
\end{equation}
in which the probability current reads 
\begin{equation}
\label{eq:current_P}
J(x,t)=-\frac{\hbar}{m}\mbox{Im}\left(\Psi(x,t) \frac{d\Psi(x,t)^*}{dx}\right) \, .
\end{equation}
Integrating the Eq.~(\ref{eq:conservation}) from $x=0$ to $x=+\infty$, at the left hand-side  
we have the derivative of the probability to have the particle in the right space $x>0$, $P_R(t)$
\begin{equation}
\label{eq:conservation_2}
\frac{d P_R(t)}{dt} = J(0,t) = 
-\frac{\hbar}{m} \mbox{Im}{\left(\Psi(x,t) \frac{d\Psi(x,t)^*}{dx} \right)}_{x=0}
 \, .
\end{equation}
Assuming $\Psi(x,0)=\phi_L(x)$ at the initial time, 
by inverting   Eqs.~(\ref{eq:psi_0}),(\ref{eq:psi_1}), we have a simple expression 
for the evolution of the wave function 
\begin{equation}
\label{eq:psi_t}
\Psi(x,t) =  
\cos(\theta) e^{-\frac{i}{\hbar} E_0 t} \psi_0(x) + \sin(\theta) e^{-\frac{i}{\hbar} E_1 t} \psi_1(x) \, .
\end{equation}
We now insert Eq.~(\ref{eq:psi_t}) into Eq.~(\ref{eq:conservation_2}) and we obtain 
\begin{eqnarray}
\label{eq:conservation_3}
\frac{d P_R(t)}{dt} &=& 
\frac{\hbar}{2m} \sin(2\theta)\sin(\Omega t) 
{\left( \psi_1 \frac{d \psi_0 }{dx} - \psi_0 \frac{d\psi_1}{dx} \right)}_{x=0} \nonumber \\
&=&
\frac{\hbar}{2m} \sin(2\theta)\sin(\Omega t) 
{\left( \phi_L \frac{d  \phi_R }{dx} - \phi_R \frac{d\phi_L }{dx} \right)}_{x=0}  
\end{eqnarray}
where we have used again   Eqs.~(\ref{eq:psi_0}),(\ref{eq:psi_1}) in the last line.
Comparing  Eq.~(\ref{eq:conservation_3}) with Eq.~(\ref{eq:evolution_Pr}) 
we obtain the relation between the tunneling amplitude and the two wave functions: 
\begin{equation}
\label{eq:nu_definition}
\nu = \frac{\hbar^2}{2m}
{\left( \phi_L \frac{d \phi_R }{dx} - \phi_R \frac{d \phi_L }{dx} \right)}_{x=0}  \, .
\end{equation}
For a symmetric potential, the left and right states are equal 
$\phi_R(x)=\phi(x)=\phi_L(-x)$ and we recover the standard formula 
for the tunneling amplitude Eq.~(\ref{eq:nu_definition_sym}).
We now introduce $x_s$, the crossing point of the left and right energies with the 
potential barrier $V(x_L)=\varepsilon_L$ and $V(x_R)=\varepsilon_R$.
Recalling Eq.~(\ref{eq:WKB_solution_2}), the  left and right 
wave functions $(s=L,R)$ inside the barrier are given by
\begin{equation}
\label{eq:WKB_solution_LR}
\phi_s(x) =  \sqrt{ \frac{m \omega_s}{2\pi e k_s(x)} } \times 
\left\{
\begin{array}{c}
e^{-\frac{1}{\hbar} \int^{x_L}_{x} \!  dx' k_L(x')}  \quad (x < x_L)\\
e^{-\frac{1}{\hbar} \int^{x}_{x_R} \!  dx' k_R (x')} \quad (x > x_R)
\end{array}
\right. 
\end{equation}
where $k_{s}(x)={[ 2m(V(x)-\varepsilon_{s}) ]}^{1/2}$.
Owing to our choice of the maximum's position, 
we have $dV(x)/dx=0$ at $x=0$ so that the derivative $d\phi_s/dx$ is related 
to the function itself $\phi_s(0)$, e.g. Eq.~(\ref{eq:derivative_0}).
Then the tunneling amplitude Eq.~(\ref{eq:nu_definition}) reads 
\begin{eqnarray}
\nu &=&\!\!\! \frac{\hbar}{2m}
\left( k_{R}(0) \phi_L(0) \phi_R(0) + k_{L}(0)\phi_R(0)  \phi_L(0) \right)  \nonumber \\ 
&=&\!\!\!
\frac{\hbar}{2m} 
\left( \! \sqrt{\frac{k_{R}(0)}{k_{L}(0)}} 
\! + \! \sqrt{\frac{k_{L}(0)}{k_{R}(0)}}  \! \right) \! 
\sqrt{k_{R}(0) k_{L}(0) } \phi_L(0) \phi_R(0) \nonumber \\ 
&=&\!\!\! \frac{1}{2}
\left( \sqrt{\frac{k_{R}(0)}{k_{L}(0)}} 
+  \sqrt{\frac{k_{L}(0)}{k_{R}(0)}}   \right) \,  
\sqrt{ \nu_R \,  \nu_L } \, . \label{eq:final} 
\end{eqnarray}
In the last equation, we have used the formula Eq.~(\ref{eq:t_WKB}) for the tunneling 
amplitude $\nu_{L}$ and $\nu_{R}$ for two {\sl symmetric} potentials 
defined by $V_L(x)\equiv V(x)$ for $x<0$ and $V_R(x)\equiv V(x)$ for $x>0$. 
The Eq.~(\ref{eq:final}) is the result 
Eqs.~(\ref{eq:result}),(\ref{eq:result_1}) as $\kappa_s(0)=\sqrt{2m(V(0)-\varepsilon_s)}$.

\section{Applications}
\label{sec:Applications}

\subsection{Asymmetric quartic potential}
\label{subsec:V4}
The first example of  asymmetric potential is the   
quartic potential including a linear term
\begin{equation}
\label{eq:V_4}
V(x) = 
V_0 \left\{
{\left[{\left( \frac{x}{a} \right)}^2 - 1 \right]}^2 
- 1
- \eta  \left( \frac{x}{a} \right) \right\} ,
\,\,
V_0 = \frac{m \omega^2 a^2}{8} 
\, .
\end{equation}
The positions of the minima for $\eta=0$ are at $\pm a$, 
the harmonic frequency at the bottom of the wells  is $\omega$.
For the symmetric case, the tunnel amplitude reads
\begin{equation}
\label{eq:vs_V4}
\nu = \, 4  {\left(\frac{2}{\pi}\right)}^{\frac{1}{2}}
\sqrt{  (\hbar\omega) V_0} \,\,  
e^{-\frac{16}{3} \frac{V_0}{\hbar\omega}}
 \, .
\end{equation}
When the asymmetry parameter is introduced $\eta>0$, 
a liner term is added in the potential which also 
removes  the energy degeneracy  
(see Fig.~\ref{fig:asymmetry}a). 
For $0 < \eta < 8/(3\sqrt{3})$ 
the potential is has still a maximum and two minima which 
are now shifted from $x=0$ and $x=\pm a$.
They are given by the formula 
\begin{equation}
\frac{x_n}{a} = \frac{2}{\sqrt{3}}  
\cos\left[ \frac{2\pi}{3} n + \frac{1}{3} \arctan  
\sqrt{  {\left( \frac{8}{3\sqrt{3}\eta} \right)}^2 -1}
  \right] \, ,
\end{equation}
with $n=2$ for the maximum $a_C$ and $n=1$ and $n=0$ for 
the left $a_L$ and right minimum $a_R$.
The harmonic frequency are given by 
${(\omega_s/\omega)}^2= (3/2){(a_s/a)}^2  - 1/2$ with $s=L,R$. 

After changing the origin of the $x$ axis, $x'=x-a_C$, 
and  knowing of the position of the two minima, we can use 
the formulas Eqs.~(\ref{eq:result}),(\ref{eq:result_1}). 
The prefactor Eq.~(\ref{eq:result_1}) can be easily calculated by observing 
that it corresponds to $A=(1/2)[ {(\zeta_L/\zeta_R)}^{1/4} +  {(\zeta_R/\zeta_L)}^{1/4}]$
with $\zeta_s = V(a_C)-V(a_s)-\hbar\omega_s/2 $ for $s=L,R$. 
We   use the formula for the tunneling amplitude 
in the symmetric double-well 
Eqs.~(\ref{eq:nu_s_instanton}),(\ref{eq:C_instanton}) 
to obtain $\nu_L$ for  and $\nu_R$ from the two potential $V_L(x)=V(x)$ for $x<a_c$ 
and $V_R(x)=V(x)$ for $x>a_c$

\begin{figure}[htbp]
\includegraphics[scale=0.3,angle=270.]{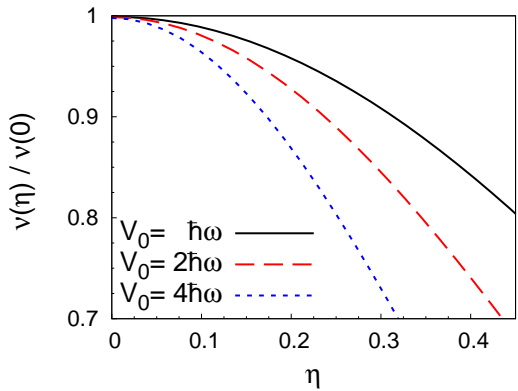}
\caption{(Color on line) The  tunneling amplitude $\nu(\eta)$ 
for the asymmetric quartic potential Eq.~(\ref{eq:V_4})   
as a function of the asymmetry parameter $\eta$ scaled 
with the symmetric value $\nu(0)$ at different values 
of the ratio $V_0/\hbar\omega$.}
\label{fig:results_quartic} 
\end{figure}

The result is given in Fig.~\ref{fig:results_quartic} where 
 the ratio between the tunneling amplitude $\nu(\eta)$ 
of the asymmetric potential and the amplitude for the symmetric case $\nu(0)$ 
is shown for different values of $V_0/\hbar\omega$.

%
%
%
%
%
%
%
%
%

Here we summarize the results.
First,  we observe that  the asymmetry of the potential always reduces 
the tunneling amplitude for the quartic potential. 
This behavior is not universal but it depends on the specific shape 
of the potential and how it is deformed (see next section).  
Second, the resonant condition $\varepsilon_L  =\varepsilon_R + \hbar\omega_R$  is matched  
at the values $\eta=0.12,0.25,0.5$ for the ratios, respectively, $V_0/(\hbar \omega)=4,2,1$ 
shown in Fig.~\ref{fig:results_quartic}.
This condition reads   $V_0/(\hbar\omega)<1/(2\eta)$ to the leading order in $\eta$  
and for $V_0/(\hbar\omega)\gg 1$ (see below and the Appendix). 
Approaching these values, the two-level description becomes insufficient. 
Third, the renormalization of the tunneling amplitude due to the asymmetry 
increases when increasing the ratio $V_0/\hbar\omega$.
This behavior can be explained by looking at $\nu_L$ and $\nu_R$.
These two amplitudes have  expressions similar to the symmetric 
case $\nu$, Eq.~(\ref{eq:vs_V4}).
The leading dependence of $\eta$ is in the exponential term, i.e. 
the term $S$ of Eq.~(\ref{eq:nu_s_instanton}) which has a scale factor 
$V_0/\hbar\omega$.

We conclude by observing  
the energy difference $\Delta\varepsilon$ overwhelms  the corrections 
of the asymmetry in the tunneling amplitude in the Rabi frequency
$\hbar \Omega = 2 {[ {(\Delta\varepsilon(\eta)/2)}^2 + \nu^2(\eta) ]}^{1/2}$. 
The last corrections scale  as $\eta^2$ at small values of $\eta$, 
as it possible to see in Fig.~\ref{fig:results_quartic}. 
On the contrary,  the energy difference increases linearly 
with respect to the asymmetry  $\Delta\varepsilon(\eta) \simeq 
V(a_R)-V(a_L) + \hbar \omega_R - \hbar \omega_L = 
2 \eta V_0 + (3/8) \eta \hbar \omega$. 

In the next example, we will consider a case in which the asymmetry affects substantially 
the tunneling dynamics.

\subsection{Asymmetric parabolic potential}
\label{subsec:PARAB}

\begin{figure}[htbp]
\includegraphics[scale=0.3,angle=270.]{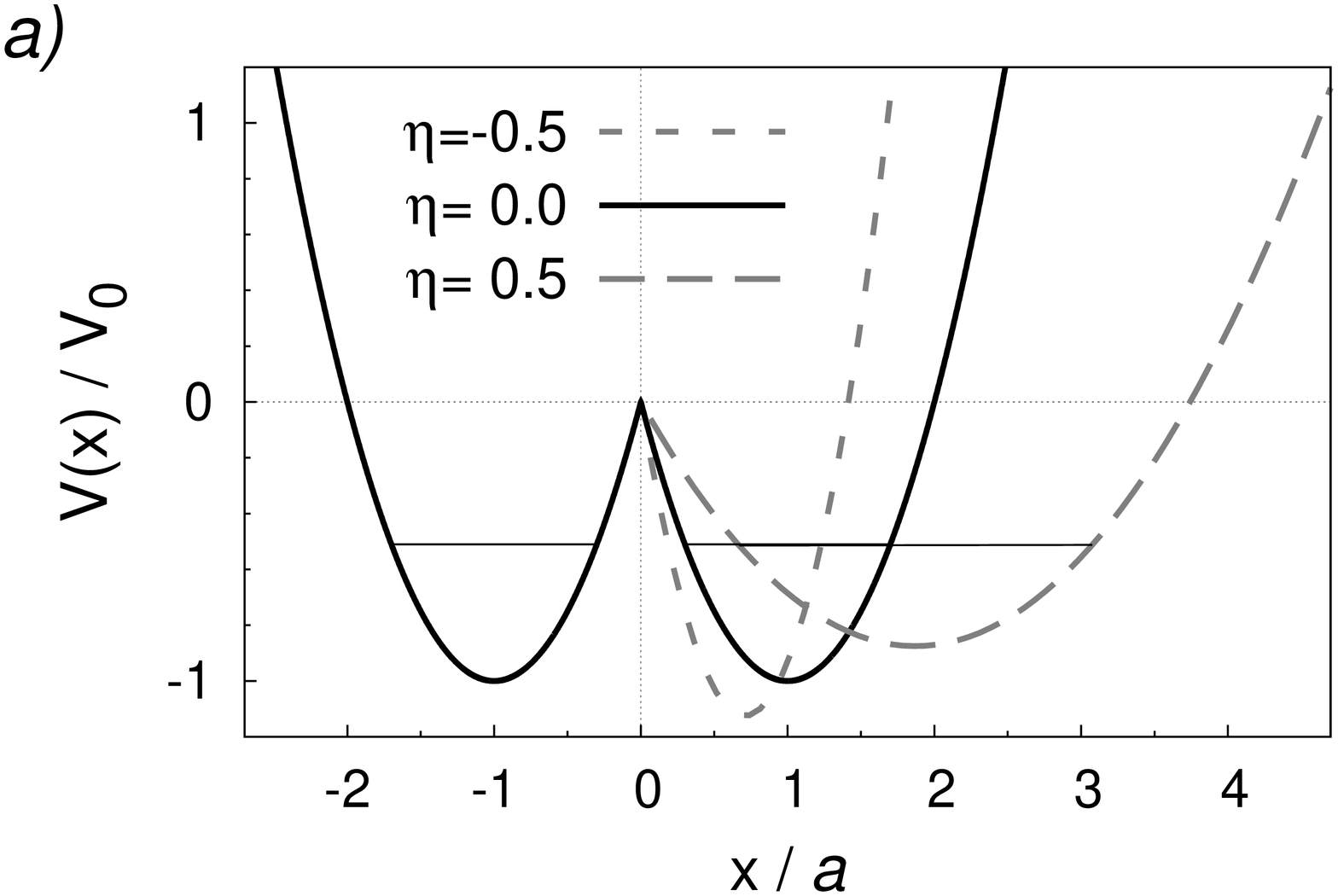}
\includegraphics[scale=0.3,angle=270.]{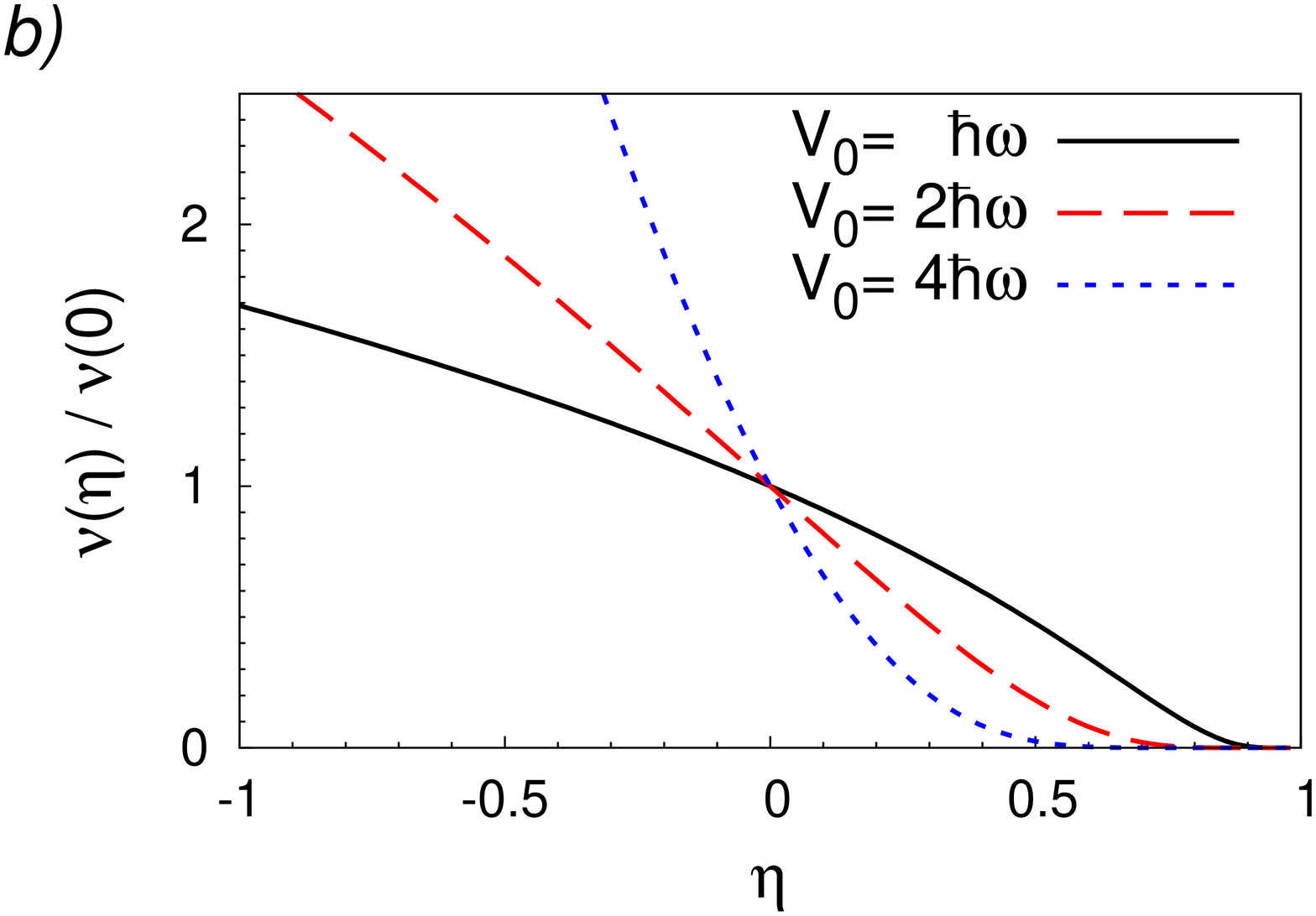}
\caption{(Color on line) a) The asymmetric parabolic potential 
Eqs.~(\ref{eq:V_asym_1}),(\ref{eq:V_asym_2}),(\ref{eq:V_asym_3}). 
b) The tunneling amplitude $\nu(\eta)$  
as a function of the asymmetry parameter $\eta$ scaled 
with the symmetric value $\nu(0)$ at different values 
of   $V_0/\hbar\omega$.}
\label{fig:results_parabolic} 
\end{figure}

The second example  belongs to the  class of asymmetric potentials 
with degenerate localized states $\varepsilon_L = \varepsilon_R$. 
An example of such potentials is the following. 
For the negative part $(x<0)$ we have $V(x)=V_L(x)$ where $V_L(x)$ reads
\begin{equation}
\label{eq:V_asym_1}
V_L(x) = V_0 
\left[
{\left( \frac{x}{a} + 1 \right)}^2 - 1 \right]
, \,\, V_0 = \frac{m \omega^2 a^2}{2}  \, .
\end{equation}
The energy of the localized states 
corresponds to $\varepsilon_L=-V_0+\hbar\omega/2$.
For the positive part $x>0$, $V(x)$ corresponds to $V(x)=V_R(x)$ 
where $V_R(x)$ reads
\begin{equation}
\label{eq:V_asym_2}
V_R(x) = V_{\eta}
\left[
{\left( \frac{x}{a_{\eta}} - 1 \right)}^2 - 1 \right]
, \,\, V_{\eta} = \frac{m \omega^2_{\eta} a^2_{\eta}}{2}  
 \, .
\end{equation}
Similarly we have $\varepsilon_R=-V_{\eta}+\hbar\omega_{\eta}/2$.
This potential is continuous  as $V_L(0)=V_R(0)$ \cite{footnote:continuity_V}.
Generally we can have $\omega_{\eta} \neq \omega $ 
and $a_{\eta} \neq a$, but the two levels are still degenerate  
under the condition
\begin{equation}
{\left( \frac{a_{\eta} \omega_{\eta}}{a\omega}  \right) }^2 
-1+\frac{\hbar\omega}{2V_0} \left( 1 - \frac{\omega_{\eta}}{\omega} \right) = 0 \, .
\end{equation}
A possible choice is 
\begin{equation}
\label{eq:V_asym_3}
\omega_{\eta}=\omega(1-\eta) , \quad 
a_{\eta}/a = \sqrt{\left(1 -\eta \frac{\hbar\omega}{2V_0} \right) }/(1-\eta) \, ,
\end{equation}
with the condition that $\eta<2V_0/(\hbar\omega)$.
The result is shown in Fig.~\ref{fig:results_parabolic}a.

Let us calculate now the tunneling amplitude $\nu(\eta)$. 
As the degeneracy is not removed $\varepsilon_R=\varepsilon_L$, the prefactor $A$
for the tunneling amplitude  in Eq.~(\ref{eq:result_1}) is one. 
In this case we have simply
\begin{equation}
\label{eq:v_as_parabolic}
\nu(\eta)=\sqrt{\nu_L \, \nu_R}=\sqrt{\nu(0) \, \nu_{R}(\eta)} \, .
\end{equation}
We have used $\nu_L=\nu(0)$ as 
the left part of the potential is unmodified (see Fig.~\ref{fig:results_parabolic}a.).
The tunneling amplitude associated to the left symmetric potential, 
$V_L(x)\equiv V(x)$ for $x<0$,  corresponds to the symmetric case with $\eta=0$: 
\begin{equation}
\label{eq:vs_parabolic}
\nu_L = \nu(0) = \, \sqrt{ \left(\frac{2}{\pi}\right)  (\hbar\omega) V_0} \, \,  
e^{-2 \frac{V_0}{\hbar\omega}} \, .
\end{equation}
As the asymmetry does not change the shape of the potential,
we can directly use the previous formula to obtain the tunneling amplitude  
associate to the right  symmetric potential, $V_R(x)\equiv V(x)$ for $x>0$, 
\begin{equation}
\label{eq:vR_parabolic}
\nu_R(\eta) = \, \sqrt{ \left(\frac{2}{\pi}\right)  (\hbar\omega_{\eta}) V_{\eta}} \, ,
e^{-2 \frac{V_{\eta}}{\hbar\omega_{\eta}}}
 \, .
\end{equation}

%
%
%
%
%
%
%
%
%

The result of Eqs. (\ref{eq:v_as_parabolic}),(\ref{eq:vs_parabolic}),(\ref{eq:vR_parabolic})
is shown in Fig.~\ref{fig:results_parabolic}b, where we report again the 
ratio between $\nu(\eta)/\nu(0)$ for different values of   $V_0/\hbar\omega$. 
We can observe that a small asymmetry of the potential $\eta<0.2$ can renormalize 
quantitatively  the amplitude $\nu$ of order $20\%$.
As  the Rabi frequency equals twice the tunneling amplitude $\hbar\Omega=2\nu(\eta)$, in this case,
the corrections due to the asymmetry appear clearly in the quantum dynamics of the 
two-level systems. 

The behavior of $\nu$ shown in Fig.~\ref{fig:results_parabolic}b can be explained 
by looking at the way  we have chosen to deform the potential Eq.~(\ref{eq:V_asym_3}). 
For the asymmetric potential, the  barrier height   at right 
 is modified as $[V(0)-V(a_R)]/V_0 = 1 - \eta \hbar\omega/(2V_0)$ 
whereas the right attempt   frequency is varied as  
$\omega_R/\omega=1-\eta$.
The second correction dominates owing to the condition   
$\hbar\omega/(2V_0) < 1$ for the two localized states.
When $\eta<0$, the right attempt frequency increases (hardening) 
leading  to an enhancement of   tunneling. 
On the contrary, for  $\eta>0$ the right attempt frequency decreases 
(softening)  leading  to a suppression of   tunneling. 
A choice different from the one of Eq.~(\ref{eq:V_asym_3}), for instance the one corresponding to 
increasing or decreasing the height of the barrier in linear way as a function of $\eta$, 
produces similar results but with a different dependence on $\eta$.

\subsection{A time-dependent problem and quantum Zeno effect}

For the asymmetric parabolic potential with degenerate local levels, 
we want to discuss a simple time-dependent problem. 

We assume to tune the asymmetry of the potential in time $\eta=\eta(t)$ so that 
the tunnel amplitude gets a time dependence $\nu= \nu(t)$. 
The time scale for the variations of  $\eta(t)$ (and therefore $\nu(t)$) 
are assumed slower enough  to avoid excitations of the systems towards higher 
energy states of the potential $E_n$ with $n\geq 2$.
Working under this assumption, the two-level description still holds. 
As a simple estimate, we consider the time scale for the variations of  $\nu(t)$ smaller 
than the harmonic frequencies $\omega_L$ and $\omega_R$ at the left and right well.
Setting the energy level to $\varepsilon=0$, we write the Schr{\"o}dinger equation 
for the two-level system as
\begin{equation}
i \hbar \frac{d}{dt}
\left(
\begin{array}{c}
 c_L(t) \\
c_R(t)
\end{array}
\right)
=\left(
\begin{array}{cc}
      0  & -\nu(t) \\
 -\nu(t) &    0 
\end{array}
\right)
\left(
\begin{array}{c}
 c_L(t) \\
 c_R(t)
\end{array}
\right) \, ,
\end{equation}
where $c_L(t)$ and $c_R(t)$ are the coefficients of the state at the time $t$.
These equations can be easily solved by introducing the sum  and 
the difference $c_{\pm}=c_L\pm c_R$ which satisfy the equation
\begin{equation}
\frac{d c_{\pm}(t)}{dt} = \pm \frac{i}{\hbar} \nu(t) c_{\pm}(t) \, ,
\end{equation}
Assuming that the system is prepared 
at the initial time $t=0$ in one localized state, let us say right, 
we have
\begin{equation}
c_L(t) = \cos\left( \frac{1}{\hbar} \int^t_0 \!\!\! dt' \nu(t')\right) , 
c_R(t) = i \sin\left( \frac{1}{\hbar} \int^t_0 \!\!\! dt' \nu(t')\right) \, , 
\end{equation}
so that the probability to remain into the initial left state reads
\begin{equation}
\label{eq:P_L_t}
P_L(t) =  {\left| c_L(t) \right|}^2 = \frac{1}{2}
\left[ 
1 + \cos\left( \frac{2}{\hbar}\int^t_0 \!\!\! dt' \nu(t')  \right)
\right]  \, .
\end{equation}
An interesting case is the following evolution for  $\nu(t)$ 
\begin{equation}
\label{eq:nu_t} 
\nu(t) = \nu_0 + \left( \nu_1 - \nu_0 \right) \sum_n \chi_n(t)  \, ,
\end{equation}
where $\chi_n(t)$ is the characteristic function equals to $\chi_n=1$ 
in the time intervals $[n(t_0+t_1) + t_0]< t  < (n+1) (t_0+t_1)$ 
and $\chi_n=0$ elsewhere.
$t_0$ is a time interval in which the tunneling amplitude is constant 
and equals to $\nu=\nu_0$ whereas in a subsequent smaller interval $t_1 < t_0$, 
the  tunneling amplitude is suppressed to $\nu_1 \ll \nu_0$ \cite{footnote_tau}. 
The signal is repeated many times during a half Rabi period $T/2 = \pi/\Omega=\pi\hbar/(2\nu_0)$.
%
%
%
%
%
%
%
\begin{figure}[hbtp]
\includegraphics[scale=0.3,angle=270.]{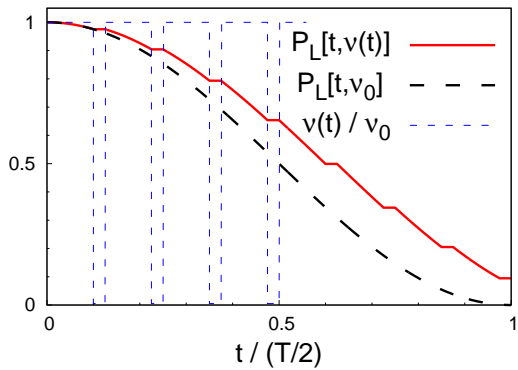} 
\caption{(Color on line)  
The probability $P_L(t)$, full (red) line, Eqs.~(\ref{eq:P_L_t}),(\ref{eq:nu_t}) 
for an asymmetric potential with degenerate local states.
Parameters:  $\nu_1/\nu_0=0.005$, $t_1/t_0=1/8$ and  $t_0= T/16$ 
where $T=\pi\hbar/(2\nu_0)$.  
The dashed (black) line is the probability $P_L$ calculated with constant amplitude $\nu_0$.
The dotted (blue) line represents the ratio $\nu(t)/\nu(0)$.}
\label{fig:Zeno} 
\end{figure}
%
%
%
%
%

The results for evolution of the probability $P_L(t)$ are shown in Fig.~\ref{fig:Zeno} both 
for the case  $\nu=\nu(t)$, Eq.~(\ref{eq:nu_t}), and for the case of constant amplitude $\nu=\nu_0$.
In  Fig.~\ref{fig:Zeno}, we can observe that the probability $P_L[t,\nu(t)]$ is always higher 
then $P_L[t,\nu_0]$. 
Reducing $\nu$ to $\nu_1$ during a short time interval $t_1$
corresponds to deform the potential in time in a way to 
suppress the tunneling amplitude $\nu_1 \ll \nu_0$.
This corresponds to trap back the particle in the starting well at regular 
time intervals. 
As a consequence, a slow-down of the probability to escape from the initial well occurs,  
a result that is referred as quantum Zeno effect in the 
literature \cite{Misra:1977,Smerzi:2006}.

\section{Conclusion}

In summary, we have derived a useful and succinct   expression 
for the tunneling amplitude $\nu$ in asymmetric double-well potentials.  
We applied it to two examples: the quartic  potential 
with a linear    force and a kind of parabolic 
potential in which the asymmetry does not remove the energy degeneracy of the two localized levels. 
From these simple examples, one can learn that there are no systematic effects 
of asymmetry on quantum tunneling. 
The tunneling amplitude is enhanced or reduced depending on the shape of the potential 
and how the asymmetry is introduced. 
However we have illustrated as the formulas Eqs.~(\ref{eq:result}),(\ref{eq:result_1}) 
allow one to obtain analytically and in a direct way the renormalization of the tunneling 
amplitude in an asymmetric double-well potential in order to discuss its behavior 
as varying the asymmetry.

\acknowledgments 

The author thanks F.W.J. Hekking and R. Whitney for useful discussions 
and critical reading of the manuscript. 
This work was supported by ANR through contracts DYCOSMA and QUANTJO. 
G.R. acknowledges support from the European networks MIDAS, SOLID and GEOMDISS.

\appendix
\section{Comparison with exact numerical results}

In this appendix we show the comparison between the exact numerical 
result of the energy splitting $(E_1-E_0)$ and the WKB semi-classical formula 
for the Rabi frequency $\hbar \Omega = 2 {[  {(\varepsilon_L(\eta)-\varepsilon_R(\eta))}^2 /4 + \nu^2(\eta) ]}^{1/2}$ 
with $\nu(\eta)$  given by the Eqs.~(\ref{eq:result}),(\ref{eq:result_1}) and the Eq.~(\ref{eq:nu_s_instanton}).

The first two eigenstates $\psi_0(x),\psi_1(x)$ and their energies $E_0,E_1$ were computed 
numerically by discretizing the time-independent Schr{\"o}dinger equation for the two 
potentials $V(x)$ discussed in the paper.
Using this approach and imposing the boundary conditions $\psi_n(x)=0$ at the end points of 
a finite interval, the Schr{\"o}dinger equation becomes a linear eigenvalue  problem 
with a tridiagonal matrix assuming that the difference between the exact value of 
the second derivative $d^2\psi(x)/dx^2$and  its discretized form 
$[\psi(x_{i+1}) +  \psi(x_{i-1}) - 2 \psi(x_{i})]/\Delta x^2$ is 
small \cite{footnote:high-E-cutoff}.
The first low-energy eigenstates are smooth functions which extend over 
lengths $\sigma_s={[  \hbar/(m\omega_s) ]}^{1/2}$.
Then  one can choose a spacing 
$\Delta x = min(\sigma_L,\sigma_R)/N$  with $N$  sufficiently large $(N \sim 10^3)$ in order 
to compute the eigenvalues with an acceptable error $\delta E_1,\delta E_0 \ll E_1-E_0$. 
The end points were set to $x_{min} \sim - M \sigma_L$ on the left, and to $x_{max}= M \sigma_R$ on the right 
$(M\sim 5-10)$. 
A scaling analysis of $E_0,E_1$ as functions of $N,M$ was also carried out to test the convergence.

\begin{figure}[hbtp]
\includegraphics[scale=0.27,angle=270.]{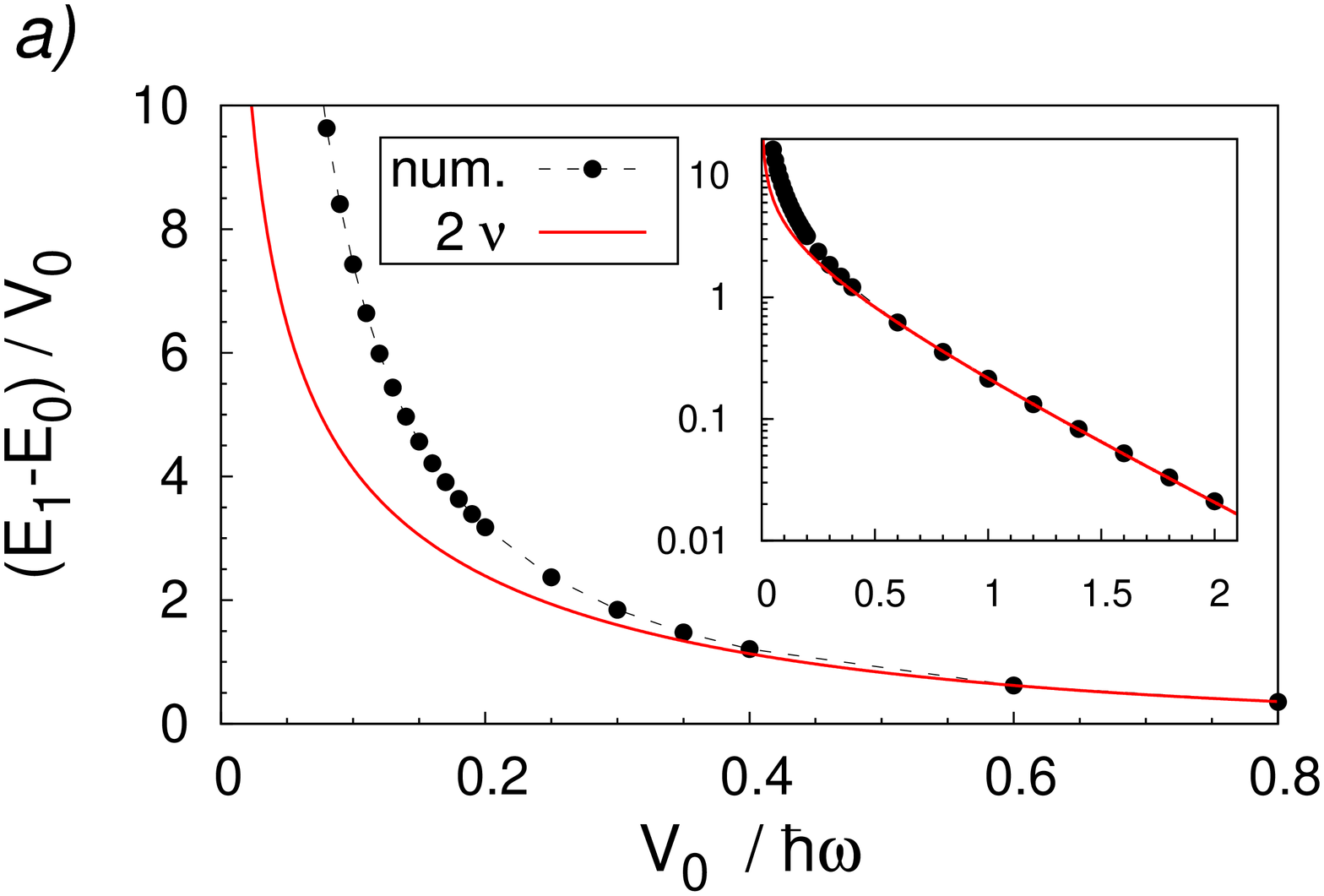}\\[-5mm]
\includegraphics[scale=0.27,angle=270.]{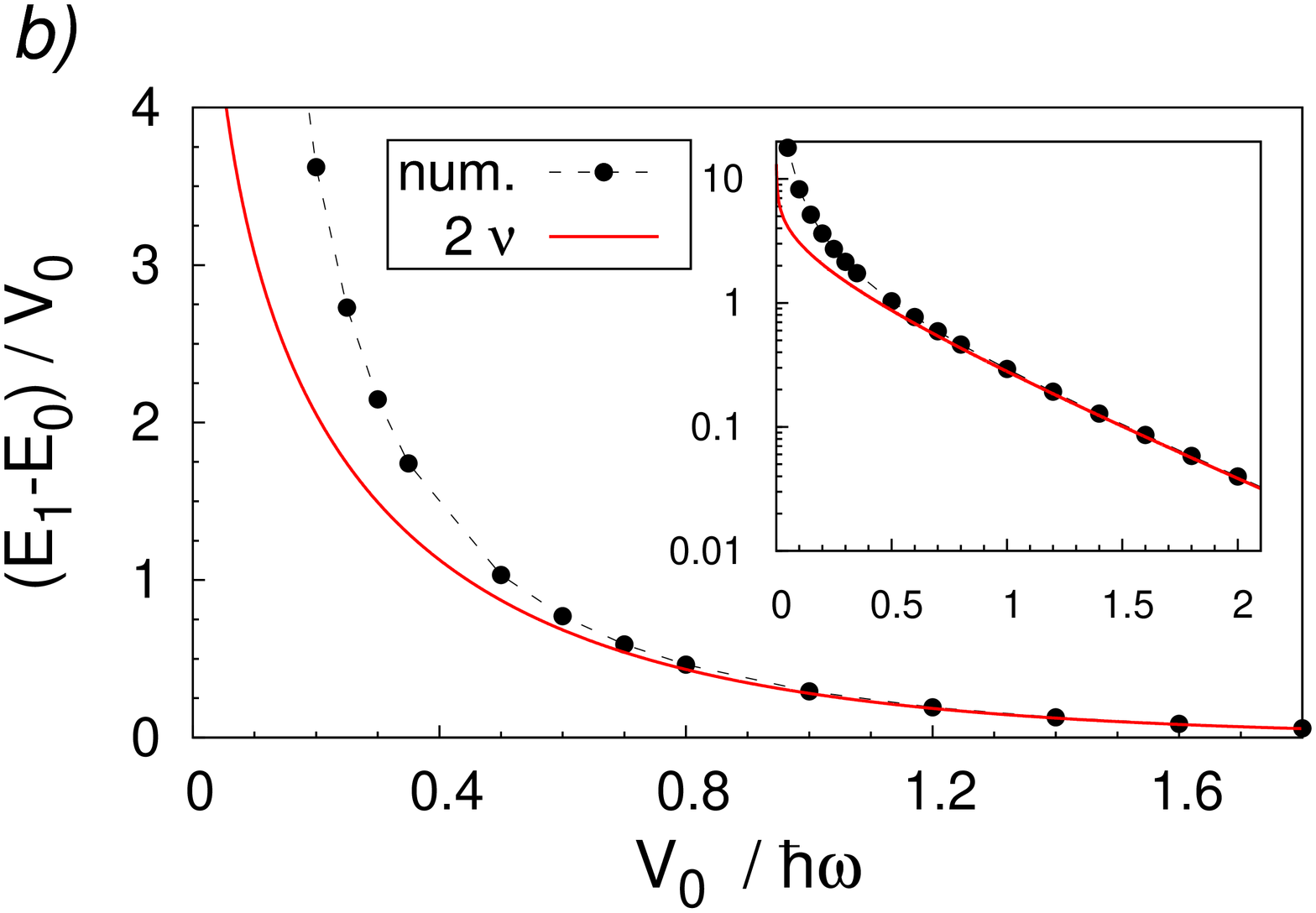}
\caption{(Color on line) Comparison between $E_1-E_0$ numerically computed 
(the black dots with dashed line) and  $2\nu$ (full red line)
for the parabolic potential. 
a) Symmetric case $\eta=0$. b) Asymmetric case $\eta=-0.5$. 
Inset: logarithmic scale for the y axis.
}
\label{fig:comparison_num_par} 
\end{figure}

In Fig.~\ref{fig:comparison_num_par} we show two examples for the double-well parabolic 
potential with two degenerate levels $\varepsilon_L=\varepsilon_R$ ($\hbar \Omega =2\nu(\eta)$), 
i.e. Eqs.~(\ref{eq:V_asym_1}),(\ref{eq:V_asym_2}),(\ref{eq:V_asym_3}) of Sec.~\ref{subsec:PARAB}.
For the symmetric case $\eta=0$ in Fig.~\ref{fig:comparison_num_par}a, 
the analytic semi-classical formula  is in agreement with the numerical result 
for  $V_0/(\hbar \omega) \agt 0.4$ (error $\alt 1\%$) 
whereas for the asymmetric case $\eta=-0.5$  in Fig.~\ref{fig:comparison_num_par}b 
the agreement is for $V_0/(\hbar \omega) \agt 0.8$. 
%
%
%
%
%
%
%

\begin{figure}[hbtp]
\includegraphics[scale=0.27,angle=270.]{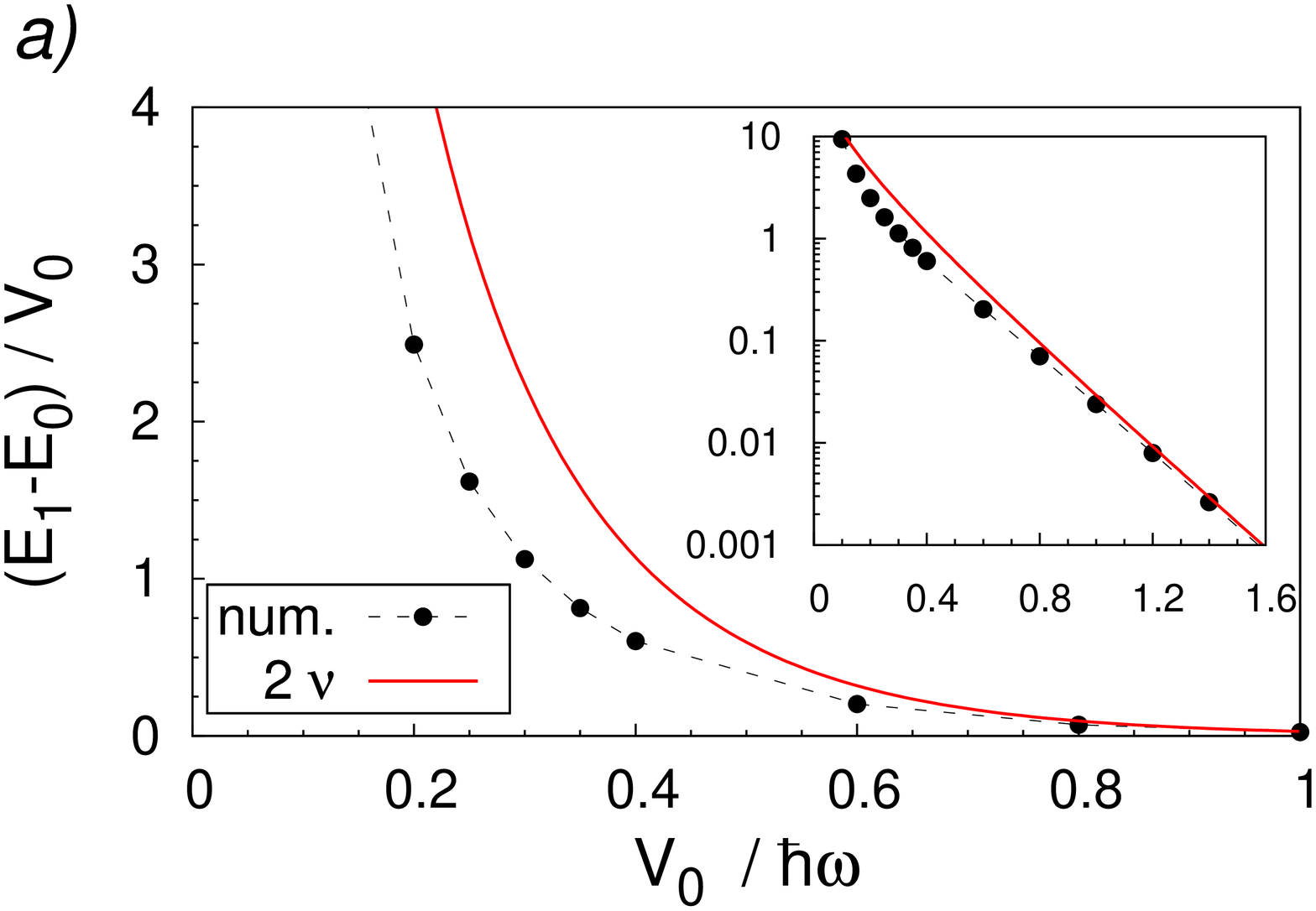}\\[-5mm]
\includegraphics[scale=0.27,angle=270.]{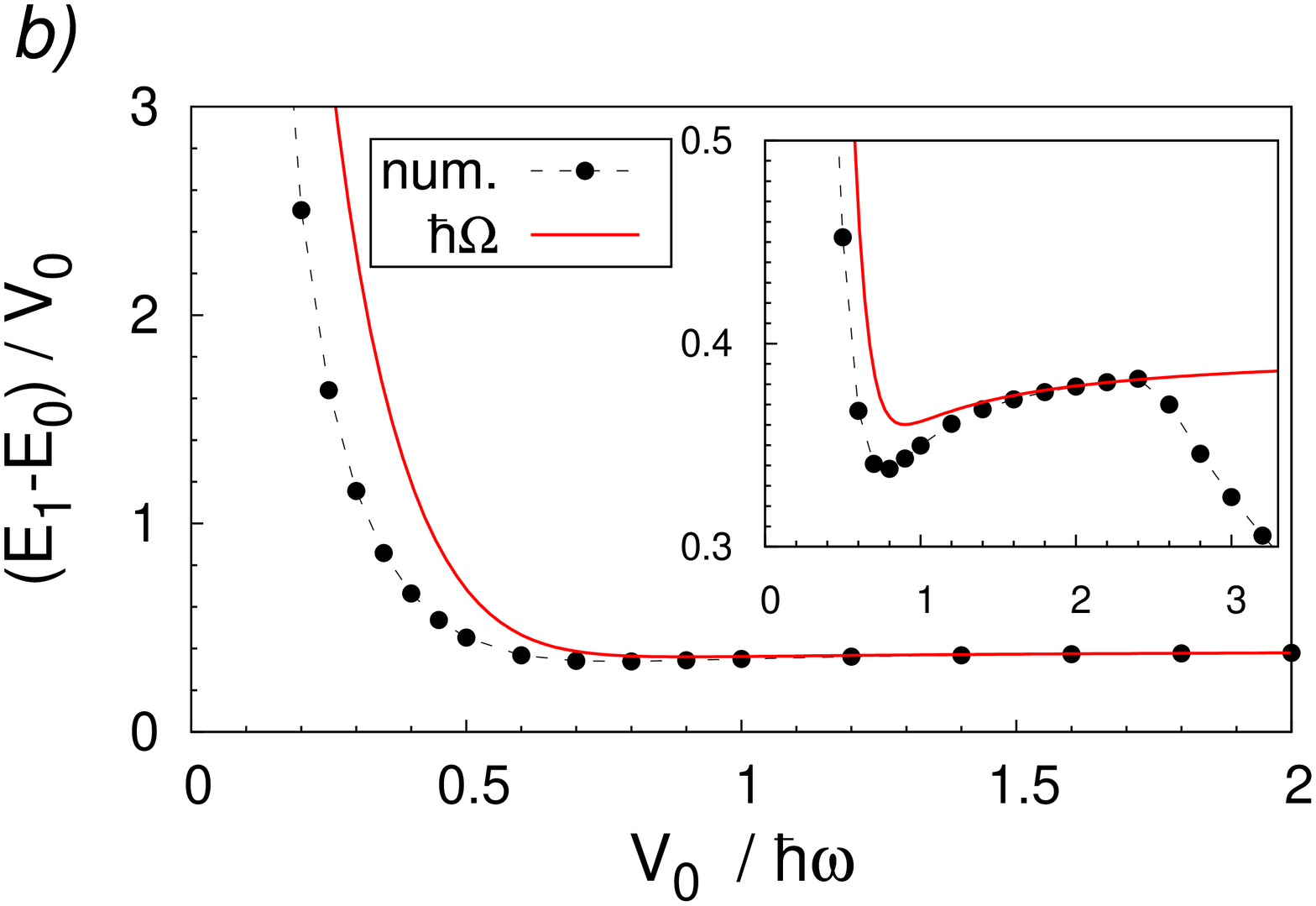} 
\caption{(Color on line) Comparison between $E_1-E_0$ numerically computed 
(the black dots with dashed line) and the semi-classical formula of the Rabi frequency (full red line) 
for the quartic potential.
a) Case $\eta=0$ ($\hbar\Omega=2\nu$).
Inset: logarithmic scale for the y axis. 
b) Case $\eta=0.2$. Inset: a close-up for the range of large ratio $V_0/\hbar\omega$. 
}
\label{fig:comparison_num_V4} 
\end{figure}

In Fig.~\ref{fig:comparison_num_V4} we compare numerical and analytic solutions 
for the bias quartic potential Eq.~(\ref{eq:V_4}) discussed in Sec.~\ref{subsec:V4}.
In this case, the asymmetry due to the linear bias $\eta>0$ removes the 
degeneracy of the localized states.
As a consequence, at given $\eta>0$, the two-level approximation breaks down 
at large values of ratio $V_0/\hbar \omega$ as approaching the resonant condition 
$\varepsilon_L = \varepsilon_R+\hbar \omega_R$ (see inset of Fig.~ \ref{fig:comparison_num_V4}b). 
Thus the upper bound of validity for our analytical approach is given by the condition
$V_0/(\hbar\omega) < (\omega_R/\omega) V_0/(\varepsilon_L-\varepsilon_R) \sim 1/(2\eta)$ 
to the leading order in $\eta$  and for $V_0/(\hbar\omega)\gg 1$.

%
%
%
%


\begin{thebibliography}{99}

%
%
%
%
%

\bibitem{Kagan-Leggett:1992} 
K.\ Kagan, A.\ J.\ Leggett, \textit{Quantum tunneling in Condensed Media} 
(Elsevier Science Publisher, 1992). 

\bibitem{Cohen-Tannoudji:1997}
C. Cohen-Tannoudji, B. Diu, F. Lalo{\"e}, \textit{Quantum Mechanics}, vol. I
(John Wiley \& Sons, 1997). 

\bibitem{Burkard:1999}
G.\ Burkard, D.\ Loss, D.\ P.\ DiVincenzo, Phys.\ Rev.\ B {\bf 59}, 2070 (1999). 


\bibitem{Levy:2007}
S.\ Levy, E.\ Lahoud, I.\ Shomroni, and J.\ Steinhauer, Nature (London) {\bf 449}, 579 (2007).

\bibitem{Anker:2005}
T.\ Anker, M.\ Albiez, R.\ Gati, S.\ Hunsmann, B.\ Eiermann, A.\ Trombettoni, 
and M.\ K.\ Oberthaler, Phys.\ Rev.\ Lett.\  {\bf 94}, 020403  (2005). 

\bibitem{Albiez:2005}
M.\ Albiez, R.\  Gati, J.\ Folling, S.\ Hunsmann, M.\ Cristiani, 
and M.\ K.\  Oberthaler, Phys.\ Rev.\ Lett.\ {\bf 95}, 010402 (2005). 

\bibitem{Shin:2005}
Y.\ Shin, G.\ B.\ Jo, M.\ Saba, T.\ A.\ Pasquini, W.\ Ketterle, and D.\ E.\ Pritchard,
Phys.\ Rev.\ Lett.\ {\bf 95}, 170402 (2005).

\bibitem{Devoret:1992}
M.\ H.\ Devoret, D.\ Esteve, C.\ Urbina, J.\ Martinis, A.\ Cleland, and J.\ Clarke, 
Chap.6 of Ref.\onlinecite{Kagan-Leggett:1992}.; 
M.\ H.\  Devoret, A.\  Wallraff, and J.\ M.\  Martinis,  arXiv:cond-mat/0411174. 

\bibitem{Makhlin:2001}
Y.\ Makhlin, G.\ Sch{\"o}n, and A.\ Shnirman, Rev.\ Mod.\ Phys.\ {\bf 73}, 357 (2001).

\bibitem{Chiorescu:2003}
I.\ Chiorescu, Y.\ Nakamura, C.\ J.\ P.\ M.\ Harmans, and E.\ Mooji, Science {\bf 299}, 1869 (2003).

\bibitem{Manucharyan:2009}
V.\ E.\ Manucharyan, J.\ Koch, L.\ G.\ Glazman, and M.\ H.\ Devoret, Science {\bf 326}, 113 (2009); 
V.\ E.\ Manucharyan, J.\ Koch,  M.\  Brink, L.\ G.\ Glazman, and M.\ H.\ Devoret, arXiv:0910.3039.

\bibitem{Simmonds:2004}
R.\ W.\ Simmonds, K.\  M.\  Lang, D.\  A.\  Hite, S.\  Nam, D.\  P.\  Pappas, 
and J.\ M.\ Martinis, Phys.\ Rev.\ Lett.\ {\bf 93}, 077003 (2004).

\bibitem{Cooper:2004} 
K.\ B.\ Cooper, M.\ Steffen, R.\ McDermott, R.\ W.\ Simmonds, S.\ Oh, D.\ A.\ Hite, 
D.\ P.\ Pappas, and J.\ M.\ Martinis,  Phys.\ Rev.\ Lett.\ {\bf 93}, 180401 (2004).

\bibitem{Johnson:2005}
P.\ R.\ Johnson, W.\ T.\ Parsons, F.\ W.\ Strauch, J.\ R.\ Anderson, A.\ J.\ Dragt, C.\ J.\ Lobb, and 
F.\ C.\ Wellstood, Phys.\ Rev.\ Lett.\ {\bf 94}, 187004 (2005).

\bibitem{Leggett:2002}
A.\ J.\ Leggett, J.\ Phys.\ Cond.\ Matt.\ {\bf 14}, R415 (2002).

\bibitem{Ladd:2010}
T.\ D.\ Ladd, F.\ Jelezko, R.\ Laflamme, Y.\ Nakamura, C.\ Monroe, and J.\ L.\ O'Brien 
Nature (London) {\bf 464}, 45 (2010)

\bibitem{Tomsovic:1998}
Tomsovic S., \textit{Tunneling in complex systems} 
(World Scientific Publishing, Singapore 1998)

\bibitem{Takagi:2002}
Takagi S., \textit{Quantum tunneling in complex systems: the semiclassical approach} 
(Cambdrige University Press, 2002)

\bibitem{Razavy:2003}
Razavy M., \textit{Quantum Theory of Tunneling} 
(World Scientific Publishing, Singapore 2003)

\bibitem{Ankerhold:2007}
Ankerhold J., \textit{Quantum tunneling in complex systems: the semiclassical approach}  
(Springer Science, Verlag-Berlin-Heidelberg 2007)

\bibitem{Miyazaki:2007}
Miyazaki T., \textit{Atom tunneling phenomena in physics, chemistry and biology} 
(Springer Science, Verlag-Berlin-Heidelberg 2004)

%
%
%
%
%

\bibitem{Dekker:1987}
H.\ Dekker, Physica {\bf 146A}, 375 (1987).

\bibitem{Schmidt:1991}
J.\ M.\ Schmidt, A.\ N.\ Cleland, and J.\ Clarke, Phys.\ Rev.\ B {\bf 43}, 229 (1991).

\bibitem{Benderskii:1999} 
V.\ A.\ Benderskii, E.\ V.\ Vetoshkin, and H.\ P.\ Trommsdorff, Chem.\ Phys.\ {\bf 244}, 299 (1999).

\bibitem{Konwent:1998}
H.\ Konwent, P.\ Machnikowski, P.\ Magnuszewski, and A.\ Radosz, J.\ Phys.\ A: Math.\ Gen.\ {\bf 31}, 7541 (1998).

\bibitem{Coleman:1977}
S.\ Coleman, Phys.\ Rev.\ D  {\bf 15}, 2929 (1977).
 
\bibitem{Kleinert:1995} 
H.\ Kleinert, \textit{Path Integral in Quantum Mechanics, Statistics and Polymer Physics}
(World Scientific, Singapore 1995, 2nd edition).

\bibitem{Garg}
A.\ Garg, Am.\ J.\ Phys.\ {\bf 68}, 430 (2000).

%
%

\bibitem{Misra:1977}
B.\ Misra and E.\ C.\ G.\ Sudarshan, J.\ Math.\ Phys.\ {\bf 18}, 756 (1977).

\bibitem{Smerzi:2006}
A.\ Smerzi, arXiv:1002.2760.

%
%
%
%
%


\bibitem{Herring:1962}
C.\ Herring, Rev.\ Mod.\ Phys.\ {\bf 34}, 632 (1962).

\bibitem{Landau:1977}
L.\ D.\ Landau, E.\ M.\ Lifshitz, \textit{Quantum Mechanics} (Pergamon, New York 1977) 3rd ed., chap. 7.

%
%
\bibitem{Gildener:1977}
E.\ Gildener, A.\ Patrascioiu, Phys.\ Rev. D {\bf 16}, 423 (1977).
%
%
\bibitem{Neuberger:1978}
H.\ Neuberger, Phys.\ Rev.\ D  {\bf 17}, 498 (1978).

%
%

\bibitem{Catelani}
A similar correction was discussed in the article 
G.\ Catelani, R.\ J.\ Schoelkopf, M.\ H.\ Devoret, and L.\ I.\ Glazman, 
Phys.\ Rev.\ B {\bf 84}, 064517 (2011).

\bibitem{Leggett:1987}
A.\ J.\ Leggett, S.\ Chakravarty, A.\ T.\ Dorsey, M.\ P.\ A.\ Fisher, 
 A.\ Garg, abd W.\ Zwerger,  Rev. Mod. Phys. {\bf 59}, 1 (1987).

\bibitem{Radosz:2006}
A.\ Radosz et al., Phys.\ Rev.\ E {\bf 73}, 026127 (2006).

\bibitem{Grabert:1985} 
H.\ Grabert, U.\ Weiss, Phys.\ Rev.\ Lett.\ {\bf 54}, 1605 (1985).

\bibitem{footnote:continuity_V}
The derivative is not continuous but it is possible to consider 
a regularization function which matches the two potentials around the origin. 
This regularization is important only in an extremely small range around $x=0$.  
It does not affect the final result for the tunneling amplitude as the last one 
depends only on integrals of the potential.

\bibitem{footnote_tau}
We are considering a square wave signal in which the typical rise time $\tau$ 
is infinite, i.e.  $\tau$ is much higher than the time scales involved in the 
two-level problem: $\nu_0,\nu_1,t_0$ and $t_1$. 
Actually, $\tau$ has an upper limit given by the adiabatic condition to avoid higher 
energy excitations in the double-well.

\bibitem{footnote:high-E-cutoff}  
At given $\Delta x$, this approximation breaks down 
for the high-energy strongly oscillatory eigenstates $\psi_n(x)$. 


\end{thebibliography}
\end{document}